\documentclass[12pt]{iopart}
\usepackage{amssymb}
\usepackage{graphicx}
\usepackage{epstopdf}
\usepackage{hyphenat}
\usepackage{stackengine}
\usepackage{multirow}
\usepackage{subfig}
\usepackage{upgreek}
\usepackage{float}
\usepackage{color}

\begin{document}

\newcommand{\summ}{\displaystyle\sum}

\title[Dynamic critical properties of non-equilibrium Potts models with absorbing states]{Dynamic critical properties of non-equilibrium Potts models with absorbing states}

\author{Ahmadreza Azizi$^1$, James Stidham$^1$, and Michel Pleimling$^{1,2}$}

\address{$^1$ Department of Physics \& Center for Soft Matter and Biological Physics, Virginia Tech, Blacksburg, Virginia 24061-0435, USA}
\address{$^2$ Academy of Integrated Science, Virginia Tech, Blacksburg, Virginia 24061-0405, USA}

\begin{abstract}
We present extensive numerical simulations of a family of non-equilibrium Potts models with absorbing states that allows for a variety of
scenarios, depending on the number of spin states and the range of the spin-spin interactions. These scenarios encompass a voter critical point,
a discontinuous transition as well as the presence of both a symmetry-breaking phase transition and an absorbing phase transition.
While we also investigate standard steady-state quantities, our 
emphasis is on time-dependent quantities that provide insights into the transient properties of the models.

\end{abstract}


\section{Introduction}\label{intro}
Non-equilibrium phase transitions, i.e. phase transitions in systems characterized by the breaking of detailed balance, show up in a variety of forms. 
Well known examples are encountered in systems with absorbing states \cite{Hinrichsen00,Odor08,Henkel08}, in driven diffusive systems \cite{Schmittmann95},
as well as in systems that, subjected to a periodic perturbation, undergo a dynamic phase transition \cite{Chakrabarti99}. Critical phenomena
far from equilibrium are much richer than their equilibrium counterparts, yielding new universality classes (directed percolation, generalized voter universality
class, etc.) as well as novel situations.

In this work we investigate the dynamic properties of two-dimensional non-equilibrium Potts models with absorbing states that undergo a variety of phase
transitions depending on model parameters such as the number of states, $q$, of the Potts spins and the range of the
spin-spin interactions. Some steady-state properties of these systems have been elucidated in the past
\cite{Lipowski02a,Lipowski02b,Droz03a,Droz03b}. For $q=2$ and only nearest-neighbor interactions the critical point, which belongs to
the generalized voter universality class, is characterized by the fact that at this point a symmetry-breaking transition between an ordered and
disordered phase takes place at the same time as an absorbing transition between an active and an absorbing phase. This voter
critical point is split into two separate phase transitions when the interaction range is extended to include up to third nearest neighbors.
At higher temperature a symmetry-breaking transition takes place that belongs to the equilibrium Ising universality class, followed
at lower temperature by an absorbing phase transition. For $q=3$, for which only the case with exclusively nearest-neighbor interactions
has been briefly studied, a discontinuous phase transition is encountered at which both the symmetry-breaking order parameter and
the absorbing order parameter show a discontinuous behavior.

In contrast to the earlier studies that mainly focused on steady-state quantities, we present in the following an investigation
of these non-equilibrium Potts models where the emphasis is on time-dependent quantities. This allows us to gain interesting
insights into the transient properties of these systems, including the aging regime. We focus on the cases with $q=2$ and $q=3$ and
consider two different ranges for the spin-spin interactions: interactions only between nearest neighbors as well as interactions
between up to third nearest neighbors. In this way we cover different scenarios, encompassing a voter critical point, a discontinuous
phase transition as well as the appearance of a symmetry-breaking phase transition and an absorbing phase transition separated by a small
temperature interval. Whereas in the past steady-state properties have been elucidated for some of these scenarios (we are
not aware of a discussion in the literature of the $q=3$ model with third-nearest-neighbor interactions though), the dynamic properties
far from stationarity,
especially at the order-disorder phase transitions, have not yet been studied systematically. Consequently, one of the goals of this
study is to gain an understanding of the relaxation processes at these different phase transitions, including critical points
belonging to the generalized voter universality class. For this we study two-time quantities like the autocorrelation and the autoresponse.

In order to check some of our conclusions we also present data for other, related, models. On the one hand we study the voter
critical point in the generalized voter model presented by Krause, B\"{o}ttcher, and Bornholdt \cite{Krause12}, as this helps to
verify whether our results for the aging properties at a voter critical point are generic. On the other hand we also simulate
the equilibrium Potts model with $q > 4$ that exhibits a first-order phase transition. Again, these simulations allow as
to make more general statements about aging quantities at a discontinuous phase transition.

In the following Section we introduce the models and discuss the different quantities that we investigate in our numerical study.
This is followed by a detailed discussion of four different cases where we vary the value of $q$ and the range of the spin-spin interactions.

\section{Models and quantities}\label{sec2}
In a series of papers \cite{Lipowski02a,Lipowski02b,Droz03a,Droz03b} Lipowski and Droz introduced a non-equilibrium model with 
$q$ absorbing states. This model, which we call LD model in the following, is in fact very similar to the 
Metropolis algorithm for the standard $q$-state equilibrium Potts model,
but with one modification in the update scheme that yields $q$ absorbing states, 
making this a non-equilibrium model. We consider the model
on the square lattice with side length $L$ where a lattice point $i$ is characterized by a variable (Potts spin) $\sigma_i$ that 
takes on the values $\sigma_i = 0, 1, \cdots, q-1$.
The energy is given by the expression
\begin{equation} \label{energy}
{\cal H} = - \frac{1}{2} \sum\limits_{i=1}^{L^2} \sum\limits_{j \in N_i} \delta\left( \sigma_i,\sigma_j \right)
\end{equation}
where $\delta(\sigma_i,\sigma_j)$ is the Kronecker delta function, with $\delta(\sigma_i,\sigma_j)=1$ if $\sigma_i=\sigma_j$ and zero
otherwise. The sum in (\ref{energy}) is over a given neighborhood $N_i$ around site $i$. We consider two different cases: the neighborhood is composed of
only the four nearest neighbors or the neighborhood is formed by the twelve nearest neighbor spins.

When simulating the equilibrium model using the Metropolis update scheme,
one randomly chooses a site $i$ and then randomly selects a possible new value for the spin variable $\sigma_i$.
This change is accepted with the probability $\mbox{min} \left[1, e^{-\Delta{\cal H}/T} \right]$ where $T$ is the temperature of the system and
$\Delta{\cal H}$ is the change of energy associated with this change of the value of $\sigma_i$. The
LD model has the same probability than the equilibrium model with the exception of the situation where all spins in $N_i$ have the
same value as $\sigma_i$. In that case the spin $\sigma_i$ is not allowed to be updated. It is this modification that breaks detailed balance
and results in a phase with $q$ absorbing states at low temperatures \cite{footnote}.

In their work Lipowski and co-workers provided some information regarding the steady-state properties of the LD model on the square lattice.
For $q=2$ and only nearest-neighbor interactions they found that at $T \approx 1.7585$ the system undergoes a continuous phase transition
from a high temperature paramagnetic phase to a low temperature phase with a double symmetric absorbing state \cite{Lipowski02a,Droz03a}.
This phase transition, at which a symmetry breaking transition between an ordered and a disordered phase takes place at the
same time as an absorbing phase transition separating an active phase from an absorbing one,
belongs to the generalized voter universality class \cite{Dornic01,Hammal05,Canet05,Vazquez08}, a non-equilibrium critical point at which the order parameter
critical exponent $\beta =0$. \footnote{The order-disorder phase transition encountered in models with
$Z_2$ symmetry and two absorbing states belongs to the generalized voter universality class.} For $q=3$ and the same neighborhood a discontinuous
phase transition, as witnessed by the discontinuous behavior of the steady-state density of active sites,
is encountered at the temperature $T \approx 1.237$ \cite{Lipowski02a}. In \cite{Droz03a} the authors observed that for $q=2$ the voter
critical point is split into two different phase transitions when extending the neighborhood to the twelve nearest neighbors, namely a
symmetry breaking phase transition belonging to the Ising universality class at high temperature, followed at a lower temperature 
by an absorbing phase transition
belonging to the Directed Percolation universality class. This splitting of the voter critical point has been
described later through Langevin equations \cite{Hammal05} and has also been observed in other microscopic models \cite{Park15,Rodrigues15}.

The second model we consider is a generalized two-state voter model discussed by Krause, B\"{o}ttcher, and Bornholdt (KBB model in the following)
\cite{Krause12}. Introducing the number of agreeing neighbors,
\begin{equation} \label{KBB}
u_i = \sum\limits_{j \in N_i} \delta(\sigma_i,\sigma_j)~,
\end{equation}
i.e. the numbers of two-state variables $\sigma_j$ in the neighborhood $N_i$ of site $i$ that have the same value as $\sigma_i$, this model is defined through
the transition probabilities $p_{(u) \longrightarrow (n_i-u)}$ for changing the value of the variable $\sigma_i$ from 0 to 1 or from 1 to 0
in case $u$ of the $n_i$ neighbors in the neighborhood $N_i$ have the same value than $\sigma_i$. For the neighborhood
that contains the nearest neighbors only (which is the case considered in the following), $n_i=4$ and the transition probabilities for the square lattice are
given by \cite{Krause12}
\begin{eqnarray}
&& p_{(4) \longrightarrow (0)} = 0, \quad p_{(0) \longrightarrow (4)}=1 \nonumber \\
&& p_{(3) \longrightarrow (1)} = \frac{1}{1+ \exp \left(2/T \right)}, \quad p_{(1) \longrightarrow (3)} = 1- p_{(3) \longrightarrow (1)} \\
&& p_{(2) \longrightarrow (2)} = 1/2 \nonumber
\end{eqnarray}
At the temperature $T=2/\ln(3) \approx 1.8205$ \cite{footnote}, these probabilities are identical to those of the linear voter model where the new
state of the variable $\sigma_i$ is provided by one of its four neighbors chosen randomly: $p_{(4) \longrightarrow (0)} = 0$, $p_{(3) \longrightarrow (1)} = 1/4$,
$p_{(2) \longrightarrow (2)} = 1/2$, $p_{(1) \longrightarrow (3)} = 3/4$, and $p_{(0) \longrightarrow (4)}=1$. The phase transition taking place
at this temperature, which separates a disordered high temperature phase from a low temperature phase with two symmetric absorbing states,
is therefore by construction in the generalized voter universality class.

For comparison and for checks we also simulated the standard equilibrium Potts model with various numbers of states $q$ larger than 4.

While our study focuses on dynamic quantities in order to elucidate the relaxation processes encountered at the different phase transitions,
we also computed some standard magnetic quantities used to locate the phase transition temperatures. Defining the quantities 
\begin{equation}
M_n = \frac{1}{\left( q-1 \right)^n} \overline{ \left( \frac{q}{N} N_m - 1 \right)^n }
\end{equation} 
where $q$ is the number of values each variable $\sigma$ can take on, $N=L^2$ is the number of lattice sites, 
and $N_m$ is the number of majority spins, we obtain the magnetization $M$,
the susceptibility $\chi$ and the Binder cumulant $U$ through the expressions
\begin{eqnarray}
M & = & M_1 \label{mag} \\
\chi & = & \frac{1}{T} \left( M_2 - M_1^2 \right) \\
U & = & 1 - \frac{M_4}{3 M_2^2} 
\end{eqnarray}
For these static quantities, $\overline{ \cdots }$ indicates a time average as well as
an ensemble average over runs with different initial conditions and different random number sequences.

Our dynamic quantities fall into two categories, those used for the investigation of relaxation and aging processes at
symmetry-breaking transitions and those studied when characterizing absorbing phase transitions. The former one include the time-dependent magnetization $M(t)$,
given by expression (\ref{mag}) with only an ensemble average over initial conditions and random number sequences, the two-time autocorrelation
function
\begin{equation} \label{corr}
C(t,s) = \frac{1}{\left( q-1 \right)} \left( \frac{q}{N} \sum\limits_{i=1}^N \left< \delta \left(\sigma_i(t), \sigma_i(s) \right) \right> - 1 \right) ~,
\end{equation}
as well as the two-time autoresponse function \cite{Lorenz07}
\begin{equation} \label{response}
\rho(t,s) = \frac{T}{q-1} \left( \frac{q}{N} \sum\limits_{i=1}^N \left< \delta \left(\sigma_i(t),r_i \right) \right> -1 \right)~,
\end{equation}
where $\left< ... \right>$ indicates an average over different realizations of the noise \cite{Henkel10}. 
Starting from a disordered initial state, we follow the standard protocol for calculating this
response by applying for the first $s$ time steps a spatially random field with amplitude $h$. 
The random field at site $i$ is given by the expression $h_i = h \, r_i$ where the
quenched random variable $r_i$ indicates the direction of the field and
takes on one of the values $0, 1, \cdots, q-1$ \cite{Lorenz07}.
After $s$ time steps the field is removed and the relaxation of the system to the steady state is monitored for times $t > s$ with the help
of the two-time autoresponse function (\ref{response}).
In order to remain in the linear response regime,
we choose the small value $h=0.05$ for the field amplitude.

In order to probe the absorbing phase transition we measure the density of active sites $\rho_a(t)$, i.e. the fraction of sites 
for which at time $t$ not all variables $\sigma$
in the neighborhood $N_i$ are in the same state than $\sigma_i$, the time-dependent survival probability $P_s(t)$ as well as the number
of flipped spins $N_f(t)$.
For the latter two quantities we prepare systems in one of the 
absorbing states, say state 0, and change the value of the variables $\sigma$ in a small area (composed of 12 spins), setting for the $q=2$ case $\sigma_j = 1$
for every site $j$ in that area, whereas for the $q=3$ case we randomly assign the value 1 or 2 to $\sigma_j$. The survival probability
is then given by the fraction of systems that at time $t$ still contain variables with values different from 0, whereas the number 
of flipped spins $N_f(t)$ is the average number of sites $k$ characterized at time $t$ by a value $\sigma_k \ne 0$.

As usual, we measure time in Monte Carlo steps, with one Monte Carlo step corresponding to $L^2$ proposed updates.

\section{The two-state model}\label{sec3}

We first investigate in the following relaxation processes and aging phenomena for the two-state systems before discussing in the next Section the
situation where variables can take on three different values. Having two-state variables interacting with their four nearest neighbors yields a
phase transition that belongs to the generalized voter universality class. Consequently, our main interest in this Section is on the dynamic
properties far from stationarity at a voter critical point. This is followed by a quick discussion of the situation for the case where the
interaction range is extended to the twelve nearest neighbors.

\subsection{Interactions with nearest neighbors only}

As already mentioned in \cite{Lipowski02a} and \cite{Krause12}, the phase transitions encountered in the LD and KBB models with only nearest-neighbor
interactions belong to the generalized voter universality class. In a series of papers de Oliveira and co-workers have investigated 
the dynamic properties far from the steady state for the non-equilibrium
linear Glauber model \cite{deOliveira02,Hase06} and a non-equilibrium linear $q$-state model \cite{Hase10} that both exhibit 
a phase transition belonging to the generalized voter universality class. The linear character of these models allows for an analytical treatment, even away from stationarity.
For the two-time autocorrelation function, Hase {\it et al.} obtain the
expression
\begin{equation} \label{hase_autocorr}
C(t,s) \sim \frac{s}{(t-s) \ln s}
\end{equation}
when starting from a disordered initial state \cite{Hase10}. 
Cast in the standard aging scaling form \cite{Henkel10}, 
\begin{equation} \label{simple_aging}
C(t,s) = s^{-b} f_C(t/s)~, 
\end{equation}
with the
scaling function $f_C(t/s) \sim \left(t/s \right)^{-\lambda/z}$ for $t/s \gg 1$, where $\lambda$ is the autocorrelation exponent and $z$ is the
dynamic exponent, this expression formally yields the exponents $b=0$ (with logarithmic correction) and $\lambda/z=1$.

\begin{figure}[ht]
\begin{center}
\includegraphics[angle=0,width=0.6\linewidth]{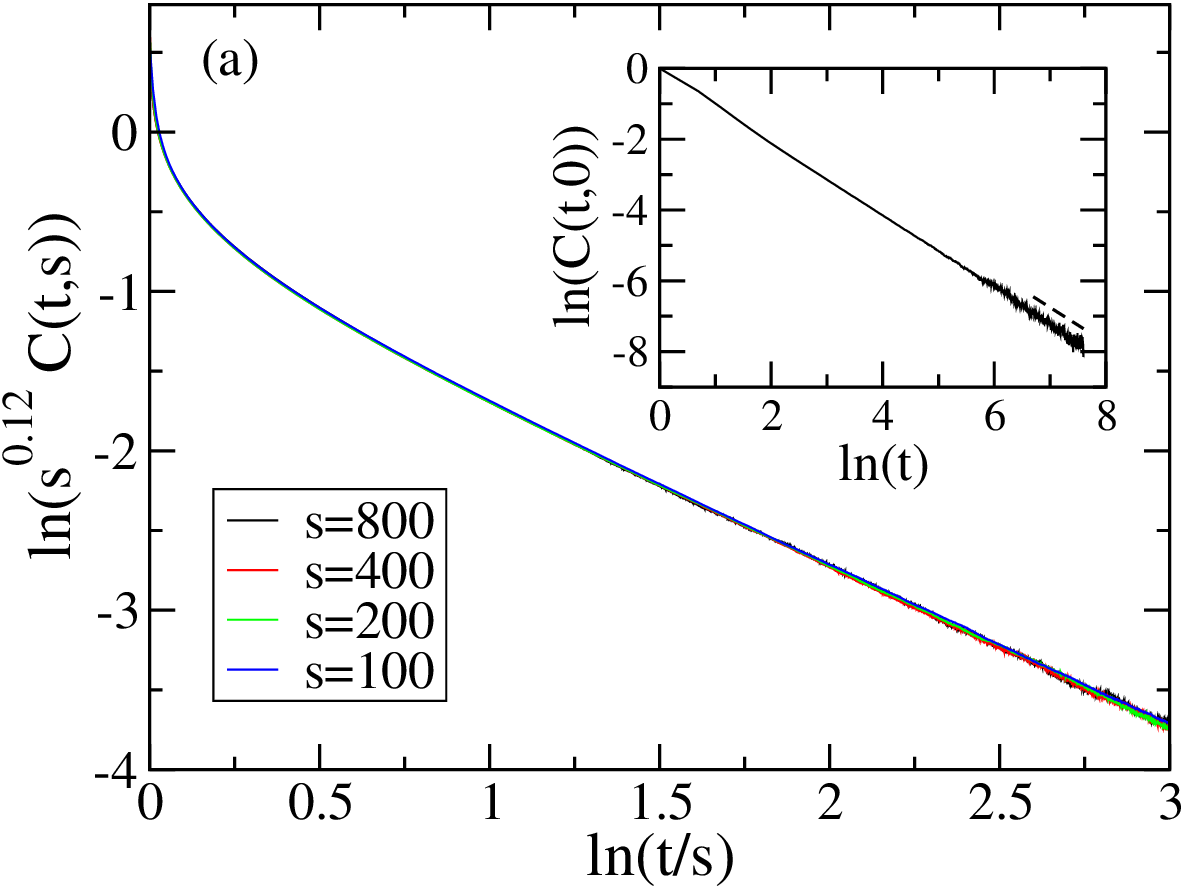}\\[0.5cm]
\includegraphics[angle=0,width=0.6\linewidth]{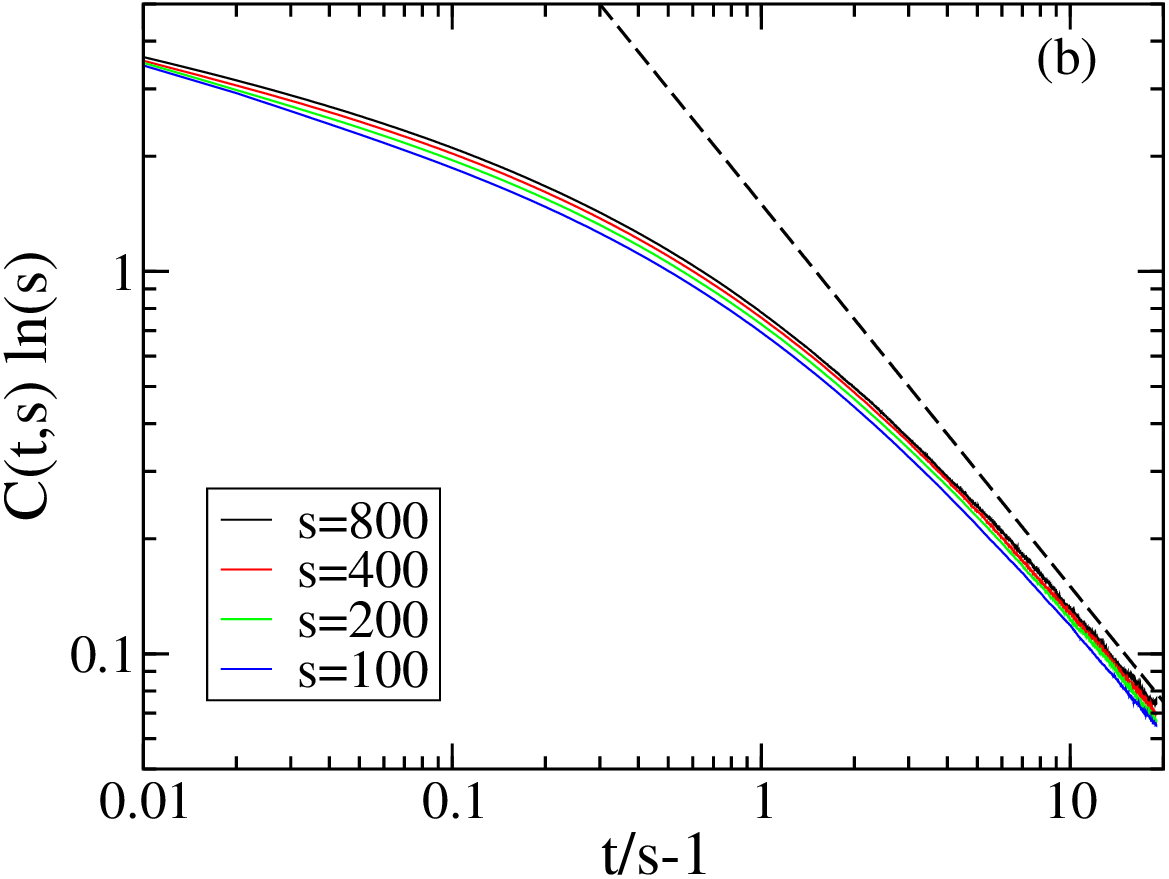}
\end{center}
\caption{(a) At the critical temperature $T_c = 2/\ln(3)$ the two-time autocorrelation function $C(t,s)$ for the KBB model shows a 
data collapse when assuming the standard aging scaling behavior (\ref{simple_aging}) with $b=0.120(5)$. 
Inset: $C(t,0)$ exhibits a power-law tail with an exponent $\lambda = 1.00(2)$ (as indicated by the dashed line). 
(b) The scaling proposed in \cite{Hase10,Chatelain11}
for the two-time autocorrelation function does not yield a data collapse nor is the expected scaling function (\ref{hase_autocorr}) (dashed line)
recovered. The data result from averaging over 8000 realizations for systems with linear size $L=800$.
}
\label{fig1}
\end{figure}

In Figure \ref{fig1} we probe the scaling properties of the two-time autocorrelation function (\ref{corr}) obtained for the KBB model. Panel (a) reveals
that a perfect data collapse is obtained when assuming the standard aging scaling (\ref{simple_aging}) with the scaling exponent $b = 0.120(5)$.
The long-time behavior of the autocorrelation is governed by a power-law decay with an exponent $\lambda/z = 1.00(2)$, see also the inset
in the figure displaying $C(t,0)$. The scaling form (\ref{hase_autocorr}) proposed in \cite{Hase10} does not result in a 
data collapse, see Figure \ref{fig1}b. In addition, the theoretical scaling function (dashed line in the figure) does not at all describe
the numerical data. In Ref. \cite{Chatelain11} the aging properties at the voter critical point were briefly discussed through a study of the
autocorrelation in a non-equilibrium symmetric three-state model at its voter point. These authors did not check for the possibility of a
standard scaling behavior, but instead exclusively plotted the data in the way shown in Figure \ref{fig1}b. 
However, comparing directly the two scenarios, as done in the two panels in Figure \ref{fig1}, reveals the superiority of the standard 
aging scaling and the discrepancy between the theoretical prediction and the numerical data.

\begin{figure}[ht]
\begin{center}
\includegraphics[angle=0,width=0.6\linewidth]{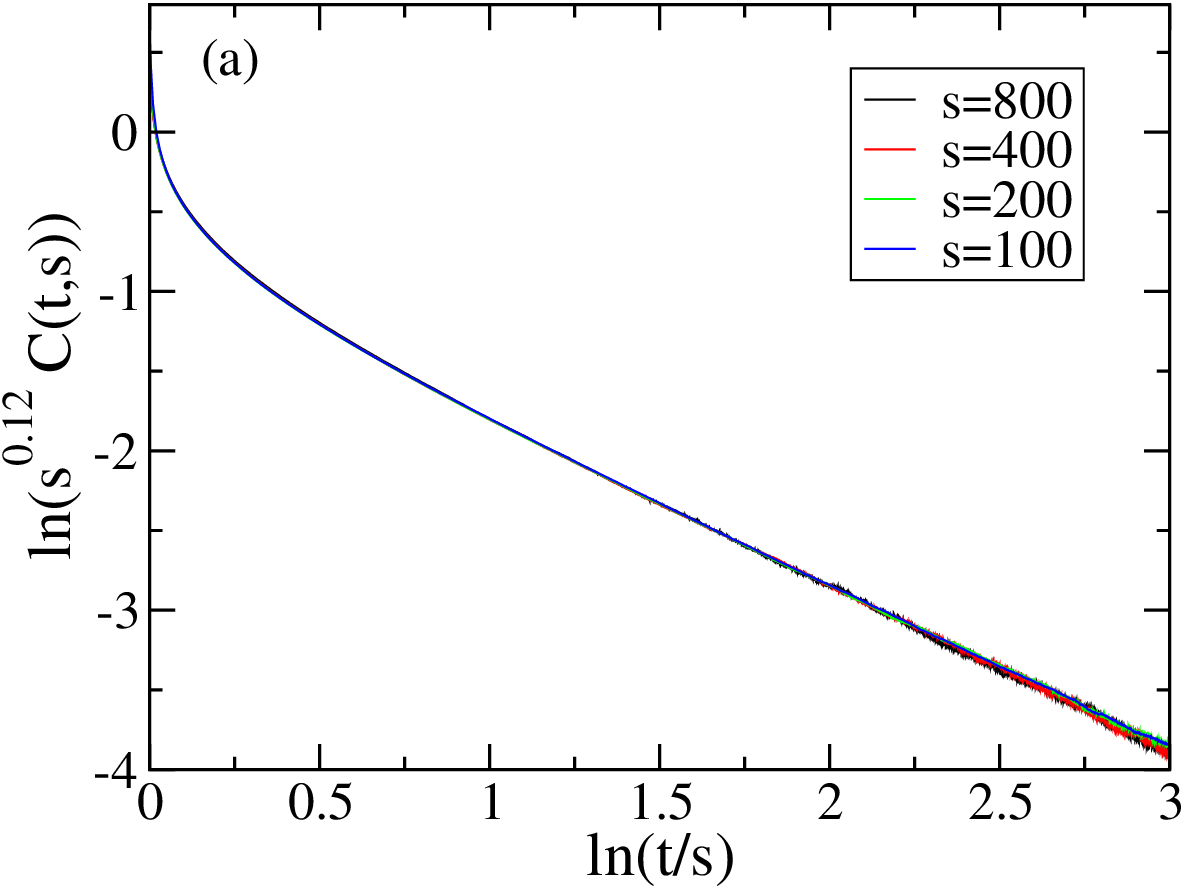}\\[0.5cm]
\includegraphics[angle=0,width=0.6\linewidth]{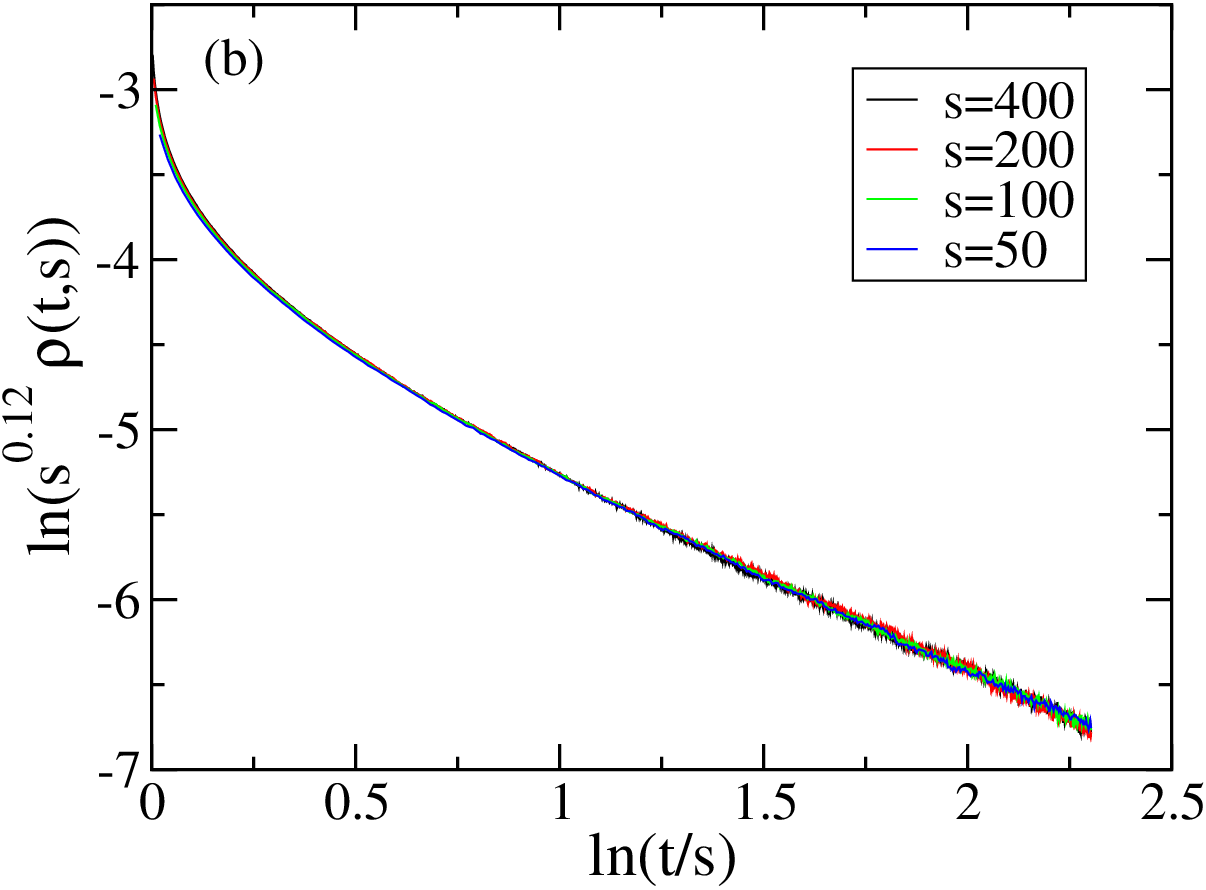}
\end{center}
\caption{(a) At the critical temperature $T_c = 1.7585$ the two-time autocorrelation function $C(t,s)$ for the LD model shows the same scaling behavior
as the critical KBB model, with the same exponents. The data result from averaging over 8000 realizations for systems with linear size $L=800$.
(b) A data collapse is also encountered for the autoresponse function $\rho(t,s)$ with the same value for the
scaling exponent. These simulations were done for systems with $400 \times 400$ sites and the amplitude $h=0.05$ for the 
spatially random magnetic field.
The data result from averaging over at least 250000 different runs.
}
\label{fig2}
\end{figure}

In Figure \ref{fig2} we show that the LD model displays the same scaling behavior as the KBB model, with the same two exponents $b=0.120(5)$
and $\lambda/z = 1.00(2)$. This simple aging scaling is encountered for both the two-time autocorrelation (\ref{corr}) and the two-time
autoresponse function (\ref{response}). We note that the autoresponse in the KBB model also displays the same scaling behavior
(not shown). These observations suggest a common, universal, behavior in the aging regime of models (both linear and non-linear) belonging
to the generalized voter universality class. This universal behavior, however, is not identical to the theoretical expression (\ref{hase_autocorr}) 
provided in \cite{Hase10}.

\begin{figure}[ht]
\begin{center}
\includegraphics[angle=0,width=0.6\linewidth]{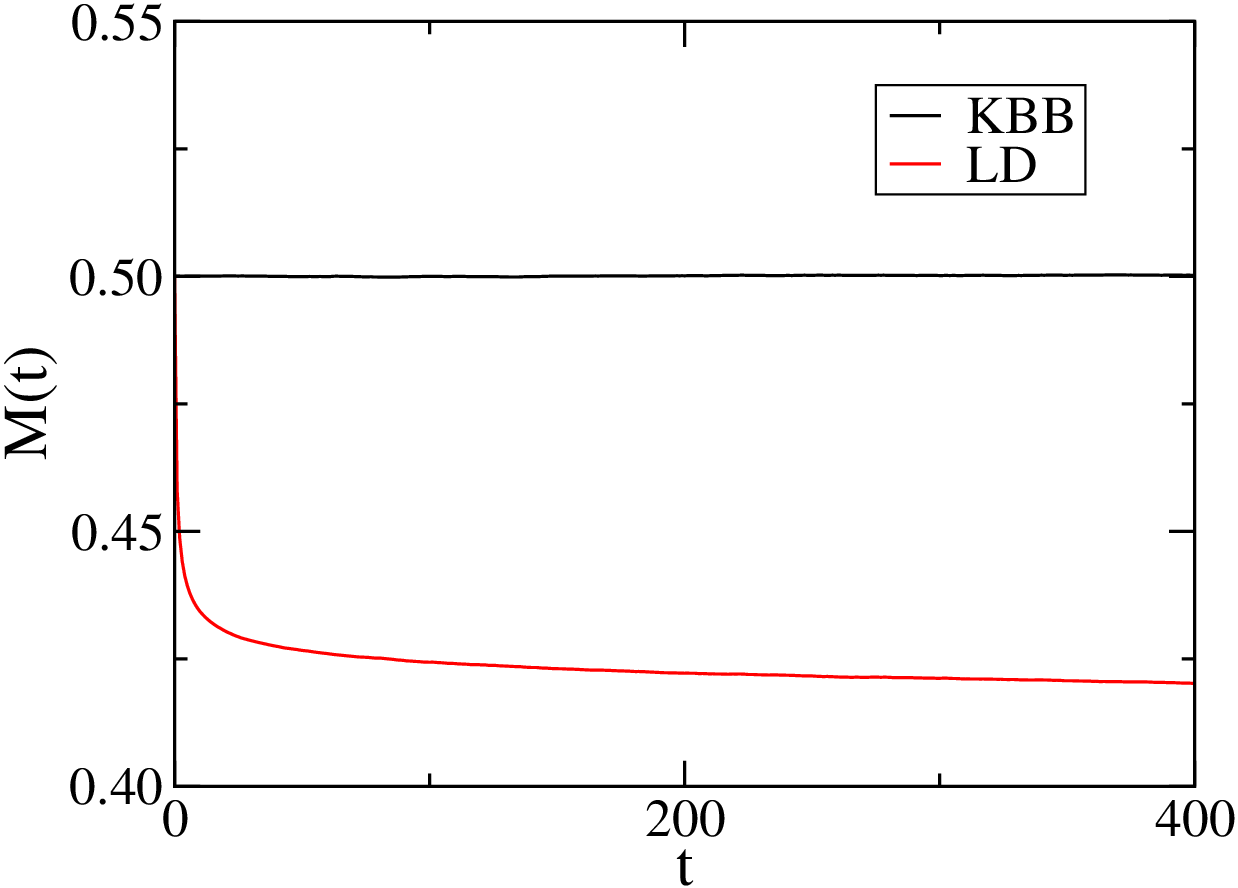}
\end{center}
\caption{Time evolution of the magnetization $M(t)$ for the critical KBB and LD models with nearest-neighbor couplings when preparing 
the system in a state with initial magnetization $M(0) = 0.5$. Systems of linear size $L=400$ have been simulated, and the data
result from averaging over 25000 independent runs.
}
\label{fig3}
\end{figure}

Whereas the linear model (we remind the reader that the KBB model at its critical point has exactly the same transition probabilities
as the original voter model) and the non-linear model display the same aging scaling for the two-time quantities, differences show
up in the time-evolution of the magnetization when starting with a magnetized (but disordered) initial state. In Figure \ref{fig3}
we show $M(t)$ when preparing the system in a state with magnetization $M(0) = 0.5$. Whereas for the KBB model the magnetization
is independent of time, $M(t) = 0.5$, in agreement with the analytical result obtained from the non-equilibrium linear $q$-state model \cite{Hase10},
for the LD model the magnetization displays a quick initial drop before slowly approaching a limit value that differs from the initial magnetization.
Our data for LD therefore agree with the earlier observation of non-conservation of the magnetization for non-linear models at their voter critical
point \cite{Castellano12}. 

\subsection{Interactions with up to third nearest neighbors}

As shown in \cite{Droz03a}, extending for the LD model the interaction neighborhood $N_i$ of a variable $\sigma_i$ to twelve sites, i.e. considering
interactions up to third nearest neighbors, splits the voter critical point into an absorbing phase transition at lower temperature 
and a symmetry-breaking phase transition at higher temperature. Based on the symmetries involved, the absorbing phase transition is expected
to belong to the (2+1) Directed Percolation universality class \cite{Hinrichsen00}, whereas the symmetry-breaking phase transition should belong to the
universality class of the two-dimensional Ising model \cite{Grinstein85}. In \cite{Droz03a} some numerical data in support of these expectations
were provided.

As already mentioned, this splitting of the voter critical point is encountered in a variety of systems, see \cite{Hammal05,Castellano09,Park15,Rodrigues15}
for some examples. Whereas it is easy to show numerically the Directed Percolation nature of the absorbing phase transition,
proving the Ising character of the symmetry-breaking phase transition is often challenging, due to the closeness of the two critical points.
These issues are well illustrated by the controversy around the phase transitions encountered in a monomer-dimer model on a square lattice
\cite{Nam11,Park12,Nam14,Park15}. 

The LD model with up to third-nearest-neighbor interactions discussed in this section is also plagued by the proximity 
of the two phase transitions. Whereas the symmetry-breaking phase transition
takes place at $T_c=4.812$, the absorbing phase transition is encountered at the slightly lower temperature of $T_a=4.738$ \cite{Droz03a}.
A direct consequence of this fact is the very narrow critical region which makes the use of standard finite-size scaling approaches challenging.

\begin{figure}[ht]
\begin{center}
\includegraphics[angle=0,width=0.6\linewidth]{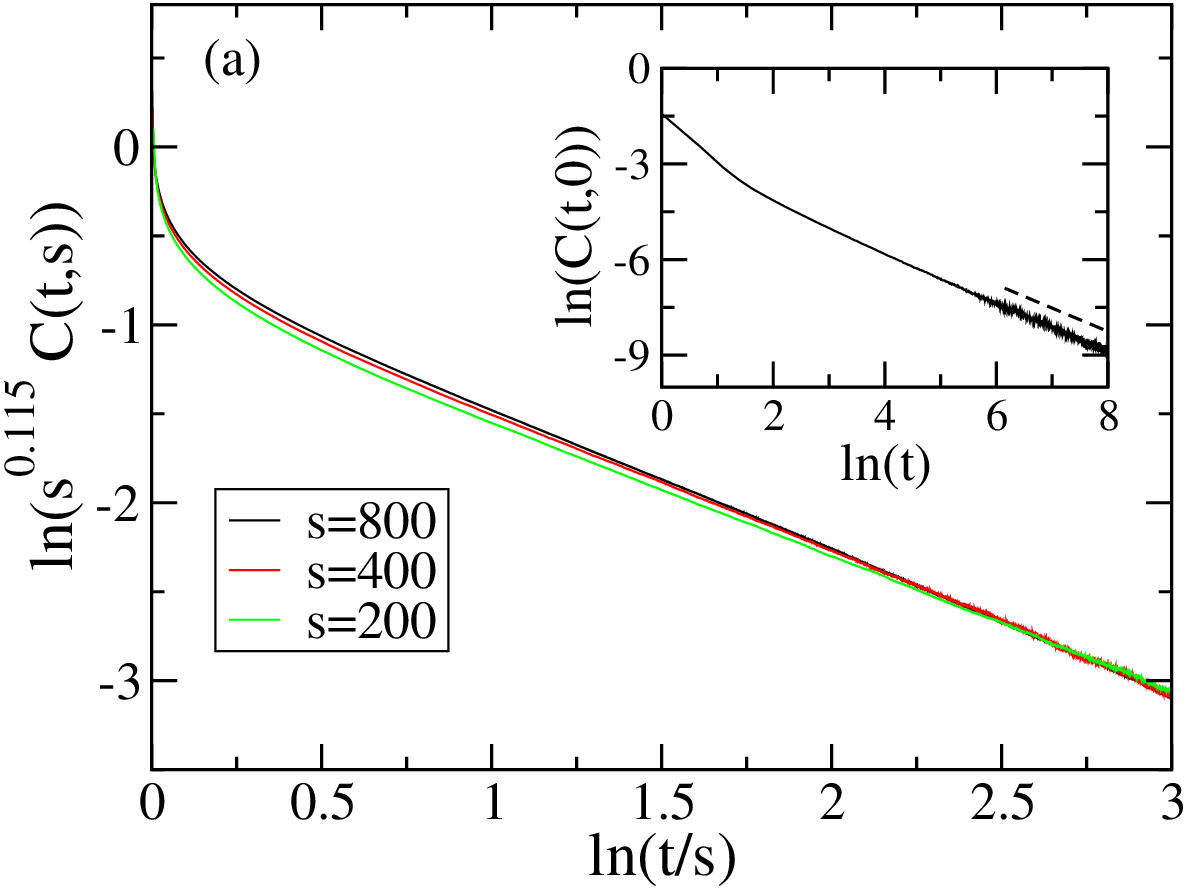}\\[0.5cm]
\includegraphics[angle=0,width=0.6\linewidth]{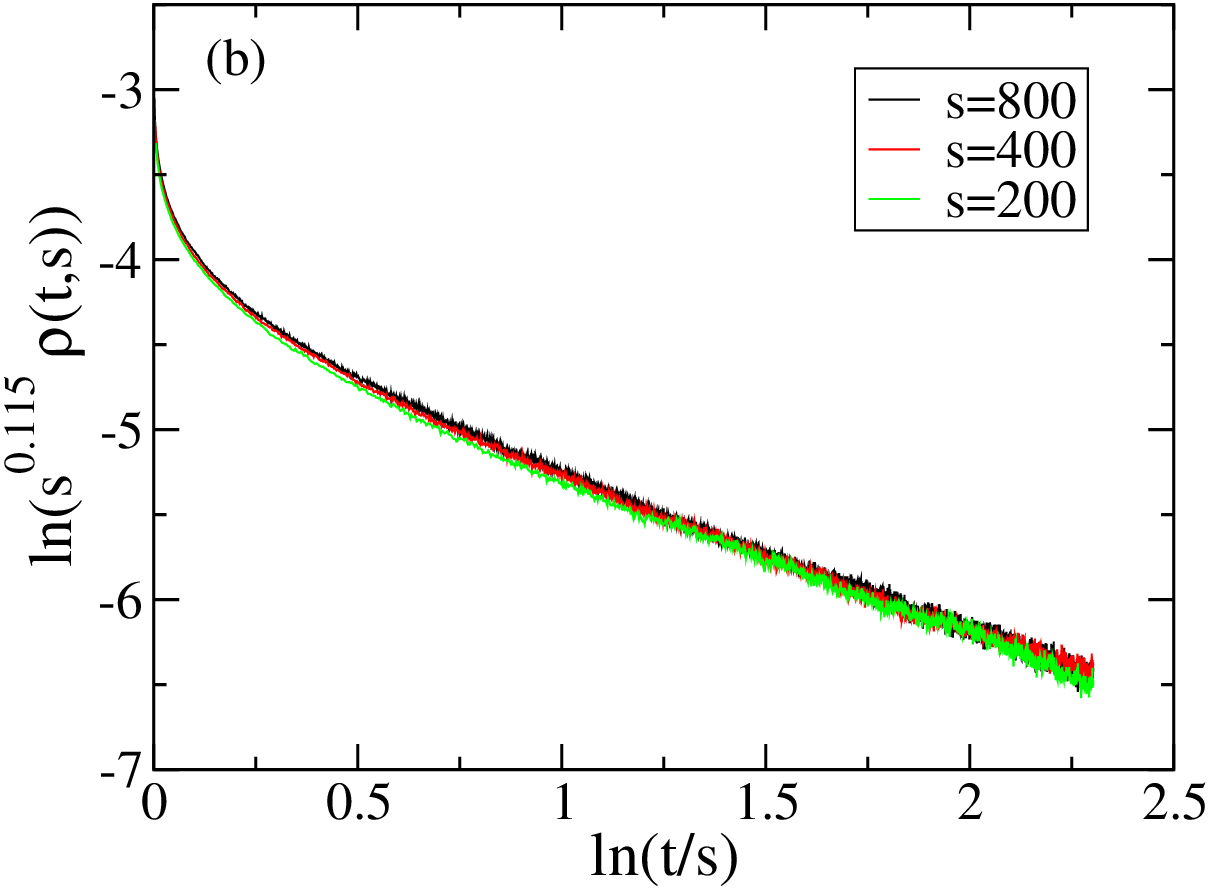}
\end{center}
\caption{(a) Scaling of the two-time autocorrelation and (b) of the two-time autoresponse function at the Ising critical point, with $T_c = 4.812$,
of the $q=2$ LD model with twelve nearest neighbors. The inset in (a) shows that the long-time decay of $C(t,0)$ is in agreement with
the known exponent $\lambda/z \approx 0.733$ (indicated by the dashed line). The autocorrelation data result from averaging over at least 10000 realizations, whereas
for the autoresponse function at least 400000 independent runs have been made. In (b) the amplitude of the magnetic field has been set to $h=0.05$. For both
quantities systems of linear extent $L=400$ have been simulated.
}
\label{fig4}
\end{figure}

In our study we focused on the question whether the measurement of dynamic quantities far from stationarity provides a viable
alternative to the standard approach that aims at determining critical exponents of static quantities. Our results are mixed. Whereas
the measurements of the initial-slip exponent \cite{Janssen89} (not shown), obtained from the increase of the magnetization when starting from a disordered initial
state with a small magnetization, or of the exponent governing the long-time power-law decay of the autocorrelation $C(t,0)$ (see inset in Figure \ref{fig4}a)
readily yield the values expected for a critical point belonging to the two-dimensional Ising universality class, the verification
of the aging scaling behavior of two-time quantities is hampered by finite-time effects that necessitate larger waiting times than usual before a clean
scaling behavior with the expected Ising exponent emerges \cite{Henkel01}. This is illustrated in Figure \ref{fig4} for both the autocorrelation (\ref{corr}) and the 
autoresponse function (\ref{response}). Plugging in the known value $2 \beta/\nu z \approx 0.115$ (where $\beta$ and $\nu$ are the usual static
exponents and $z$ is the critical dynamic exponent) for the scaling exponent,
we remark that the longest waiting times start to exhibit an acceptable data collapse. On the other hand, as shown in the inset of 
Figure \ref{fig4}a, the autocorrelation function in the long-time limit does decay with the correct exponent $\lambda/z \approx 0.733$ 
(dashed line) \cite{Pleimling05}.

Based on our data we conclude that the investigation of non-equilibrium quantities provides an alternative way of determining the universality 
class of a symmetry-breaking phase transition located in close vicinity to an absorbing phase transition, with some measurements, like that of the initial-slip
behavior of the magnetization or the long-time decay of the autocorrelation function $C(t,0)$, being more useful than others.

\section{The three-state model}\label{sec4}
We now turn to the three-state LD model with the same two interaction neighborhoods. Our goal is again to characterize the different phase transitions
through dynamic quantities measured far from stationarity.

\subsection{Interactions with nearest neighbors only}

\begin{figure}[ht]
\begin{center}
\includegraphics[angle=0,width=0.6\linewidth]{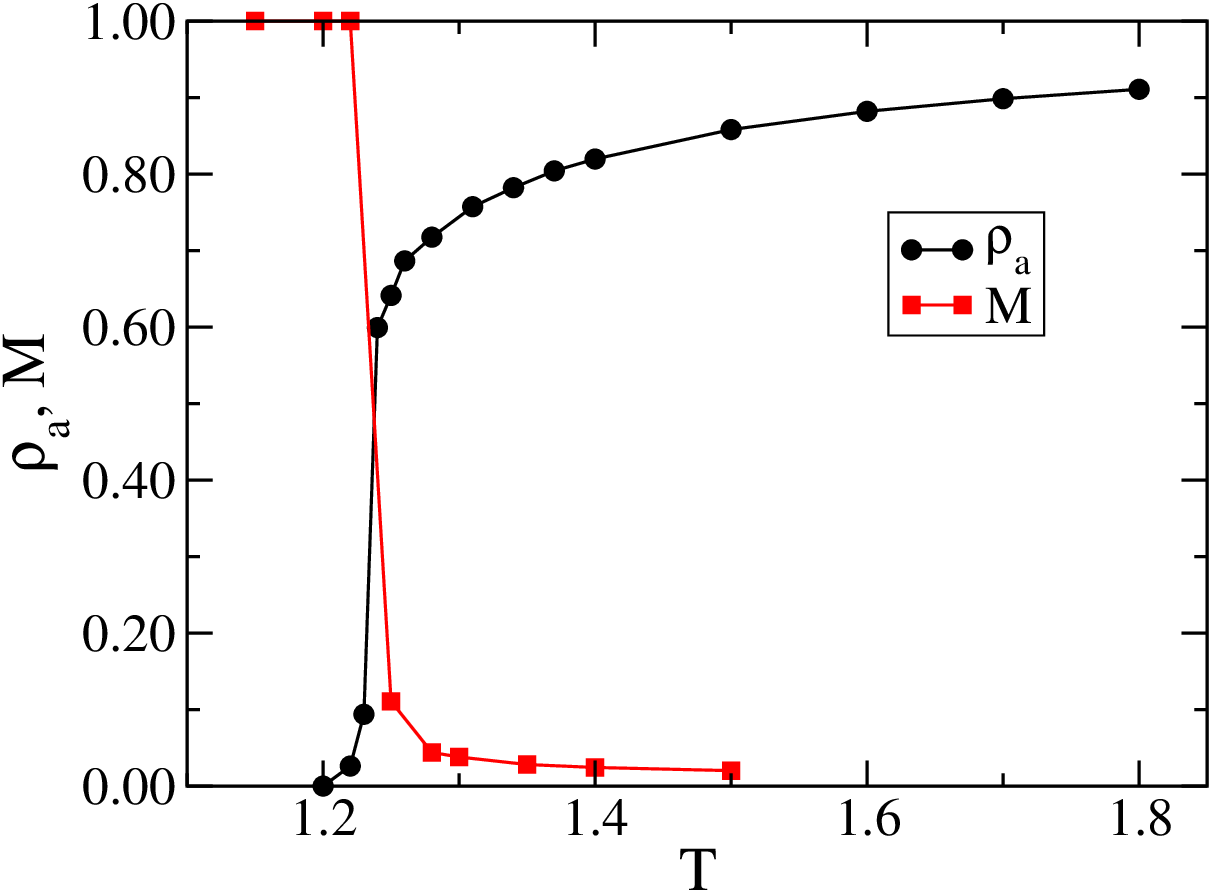}
\end{center}
\caption{The steady-state density of active sites $\rho_a$ and the steady-state magnetization $M$ as a function of temperature for the LD model with $q=3$
and only nearest-neighbor interactions. Both order parameters display a jump at the same temperature. The data, obtained for a system with
linear extent $L=400$, result from a time average (after having discarded the early time data) and an ensemble average over 100 independent runs.
Error bars are smaller than the sizes of the symbols.
}
\label{fig5}
\end{figure}

As already noted in \cite{Lipowski02a}, the two-dimensional LD model with $q=3$ undergoes a discontinuous phase transition at $T\approx 1.273$. 
As shown in Figure \ref{fig5}
the discontinuity is encountered for both the symmetry-breaking order parameter (the steady-state magnetization $M$) and the order parameter of the absorbing
phase transition (the steady-state density of active sites $\rho_a$). This is consistent with the observation \cite{Lipowski02a} that the
time-dependent survival probability displays a systematic bending without clear power-law regime.

\begin{figure}[ht]
\begin{center}
\includegraphics[angle=0,width=0.6\linewidth]{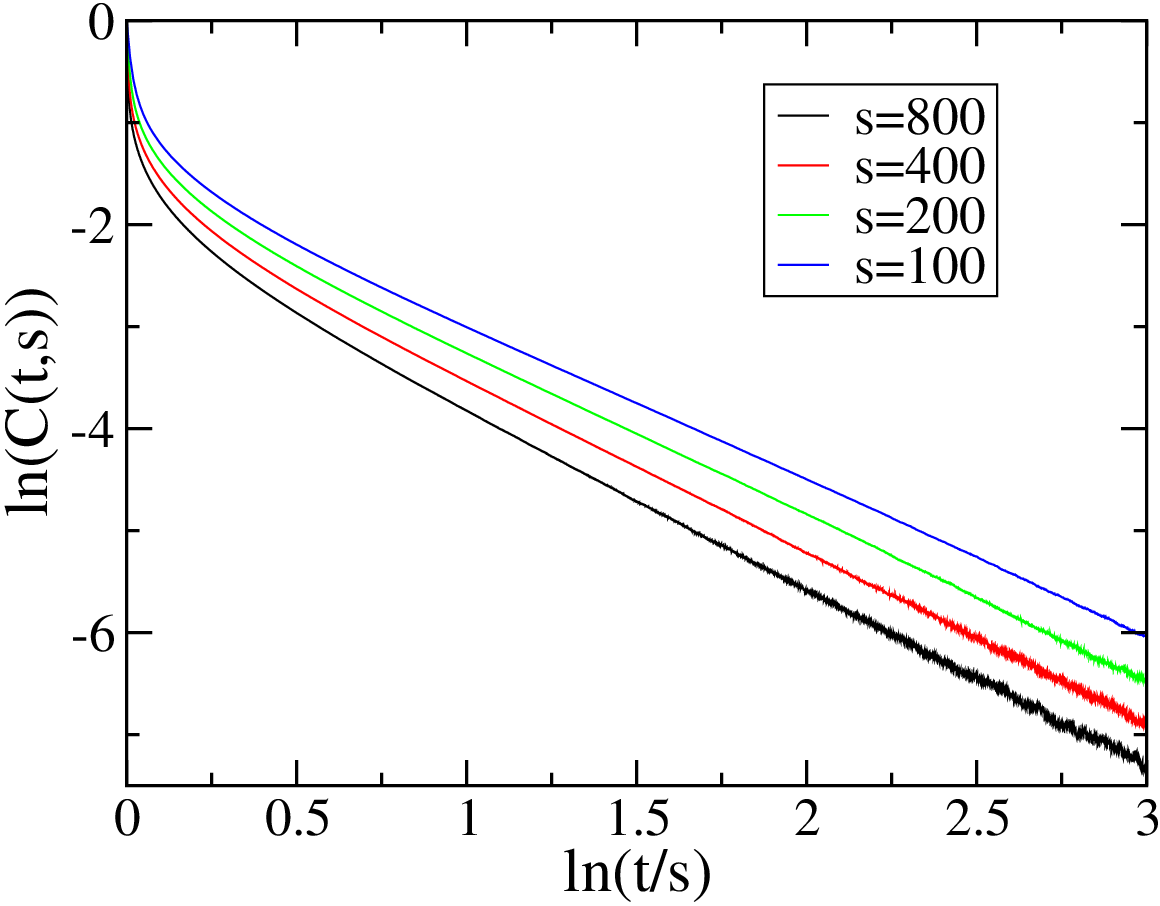}
\end{center}
\caption{Two-time autocorrelation function for the three-state LD model with only nearest-neighbor interactions at the temperature
$T=1.237$ of the discontinuous phase transition. The autocorrelation is plotted as a function of the ratio $t/s$. The linear extent of the system is $L=800$, and
10000 independent runs have been performed.
}
\label{fig6}
\end{figure}

In Figure \ref{fig6} we plot the autocorrelation function at this discontinuous phase transition as a function of $t/s$. 
At first look the autocorrelation function seems to behave in very similar ways for the different waiting times.
Close inspection of
the long time behavior, however, indicates that a fit to a power-law for the larger values of $t/s$ yields an effective exponent whose value increases
with the waiting time. Dynamical scaling therefore can not be expected for this system. 

\begin{figure}[ht]
\begin{center}
\includegraphics[angle=0,width=0.6\linewidth]{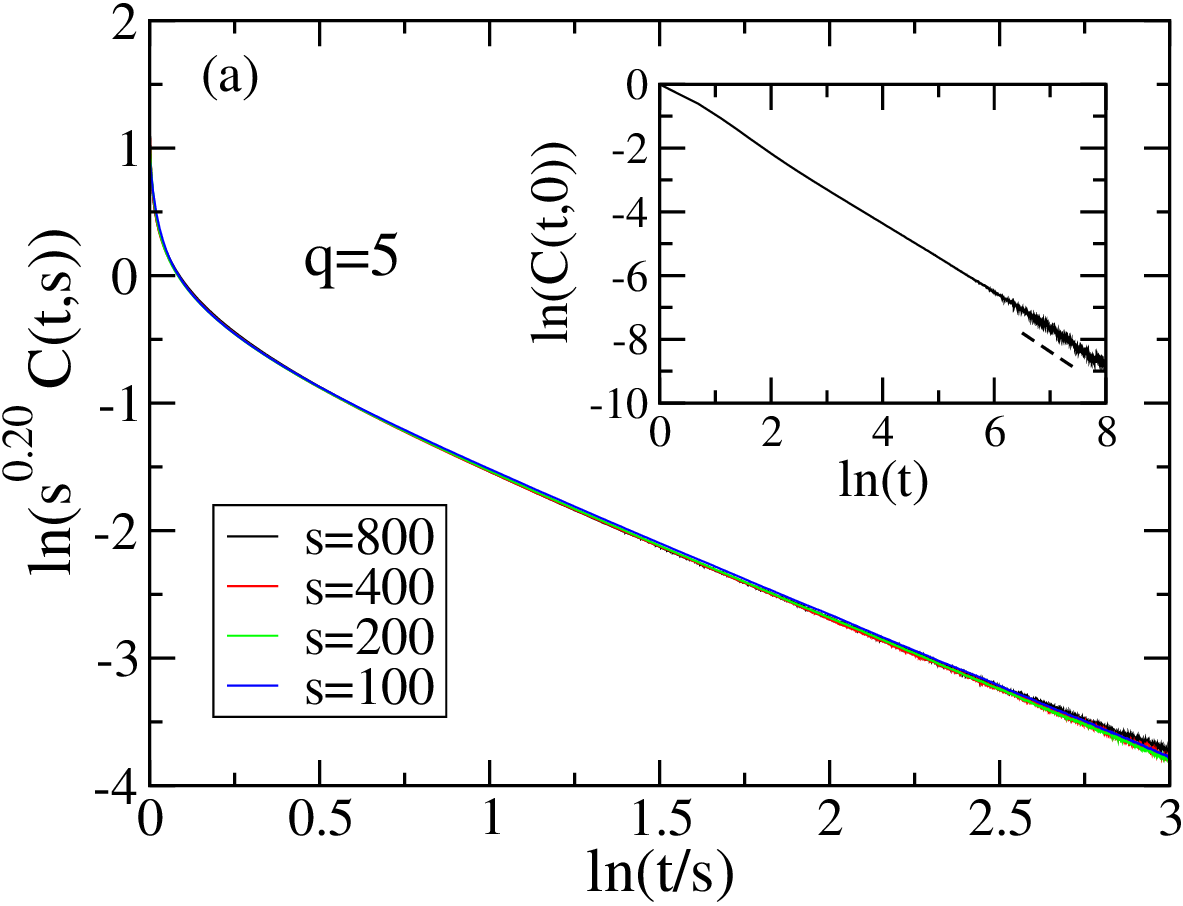}
\includegraphics[angle=0,width=0.6\linewidth]{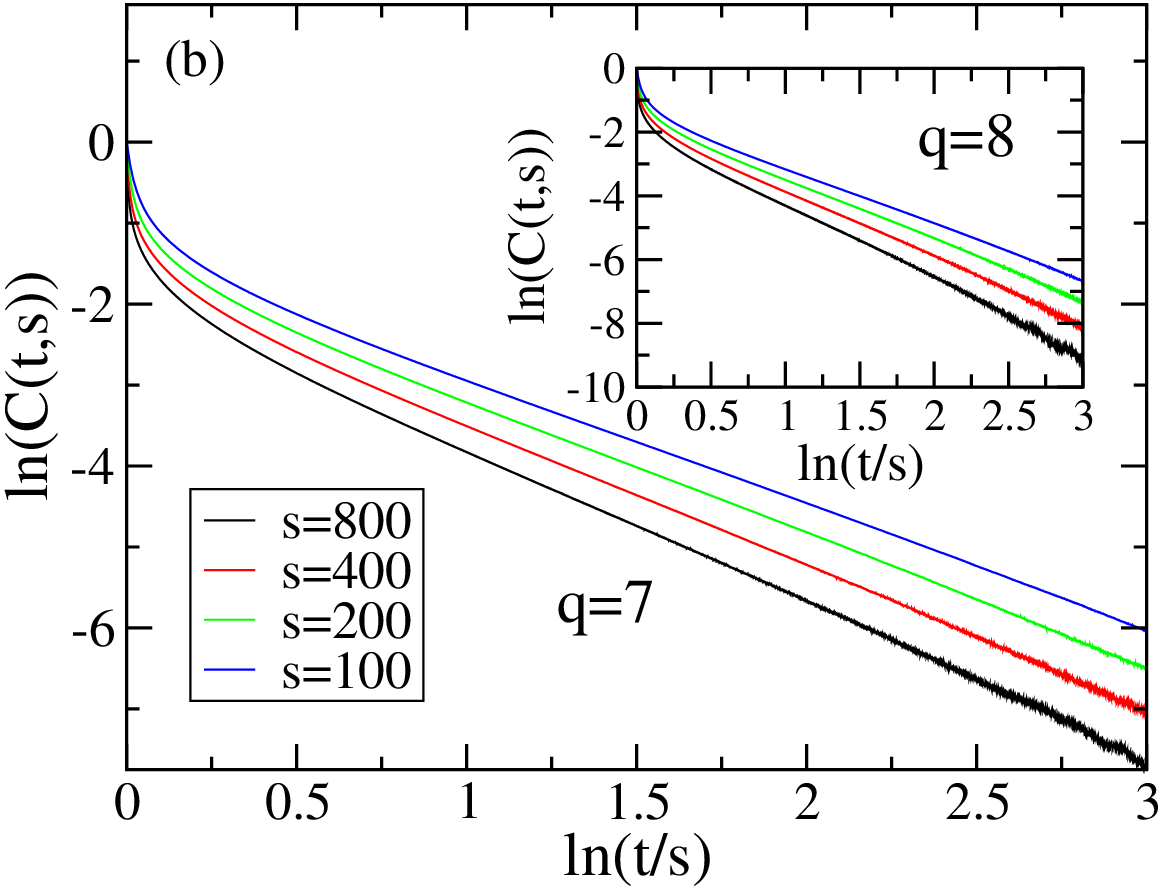}
\end{center}
\caption{Scaling of the autocorrelation $C(t,s)$ of the equilibrium Potts model with (a) $q=5$ states and (b) $q=7$ 
(inset $q=8$) states at their 
discontinuous phase transition temperature $T = \left[ \ln \left(1 + \sqrt{q} \right) \right]^{-1}$. A perfect aging scaling prevails at the
weak first-order transition for the $q=5$ case, whereas at the stronger discontinuous transitions of the $q=7$ and $q=8$ cases a simple aging data collapse 
can not be achieved. The dashed line in the inset of (a) indicates that $C(t,0)$ decays for $q=5$ algebraically with
an exponent $\sim 1.17$. The system size is $L=800$, and the data result from averaging over at least $9000$ independent
runs.
}
\label{fig7}
\end{figure}

In order to check whether this behavior is also observed at an equilibrium first-order phase transition, we extended our study to the
equilibrium two-dimensional $q$-state Potts model with $q > 4$ and computed the two-time autocorrelation at the phase transition temperature 
$T = \left[ \ln \left( 1 + \sqrt{q} \right) \right]^{-1}$. In all cases with $q > 4$ the transition is known to be of first order \cite{Wu82}.
Figure \ref{fig7} shows a remarkable difference when comparing our data for $q=5$ with those for $q=7$ and $q=8$. 
Whereas for $q=7$ and $q=8$ we have a behavior similar to that of the $q=3$ LD model, characterized by the absence of dynamical scaling, for $q=5$ we observe
a perfect aging scaling behavior with exponents $b=0.20(1)$ and $\lambda/z=1.17(2)$. In order to understand these very different results for different
first-order phase transitions, we remark that for $q=5$ we are dealing with a weak first-order phase transition characterized by a pseudocritical
behavior \cite{Peczak89}. Indeed, from the exact expression for the correlation length at the transition point \cite{Buffenoir93} one obtains
that for $q=5$ the stationary correlation length is $\xi = 2512.25$. As this is much larger than our system size, the observed relaxation processes  
do not differ from those encountered at a critical point. It is then not surprising that we observe dynamical scaling even though the system
has a discontinuous transition. For the $q=7$ respectively $q=8$ model, the value of the stationary correlation length is $\xi = 48.10$
respectively $23.88$, 
i.e. much smaller than our system size.
In this case, the dynamical correlation length can only grow up to the value of $\xi$, which results in the absence of dynamical scaling and
leads to a behavior similar to that observed in Figure \ref{fig7} for the $q=3$ LD model.

\subsection{Interactions with up to third nearest neighbors}

In this final part we discuss how the properties of the $q=3$ LD model change when we replace the interaction neighborhood $N_i$
with four sites by the neighborhood with twelve sites. The data we present in the following indicate that the single discontinuous 
transition encountered for only nearest-neighbor interactions is replaced by two continuous phase transitions when extending the
interaction range to third nearest neighbors: a symmetry-breaking phase transition belonging to the universality class of the 
two-dimensional $q=3$ Potts model and an absorbing phase transition belonging to the Directed Percolation universality class.

\begin{figure}[ht]
\begin{center}
\includegraphics[angle=0,width=0.6\linewidth]{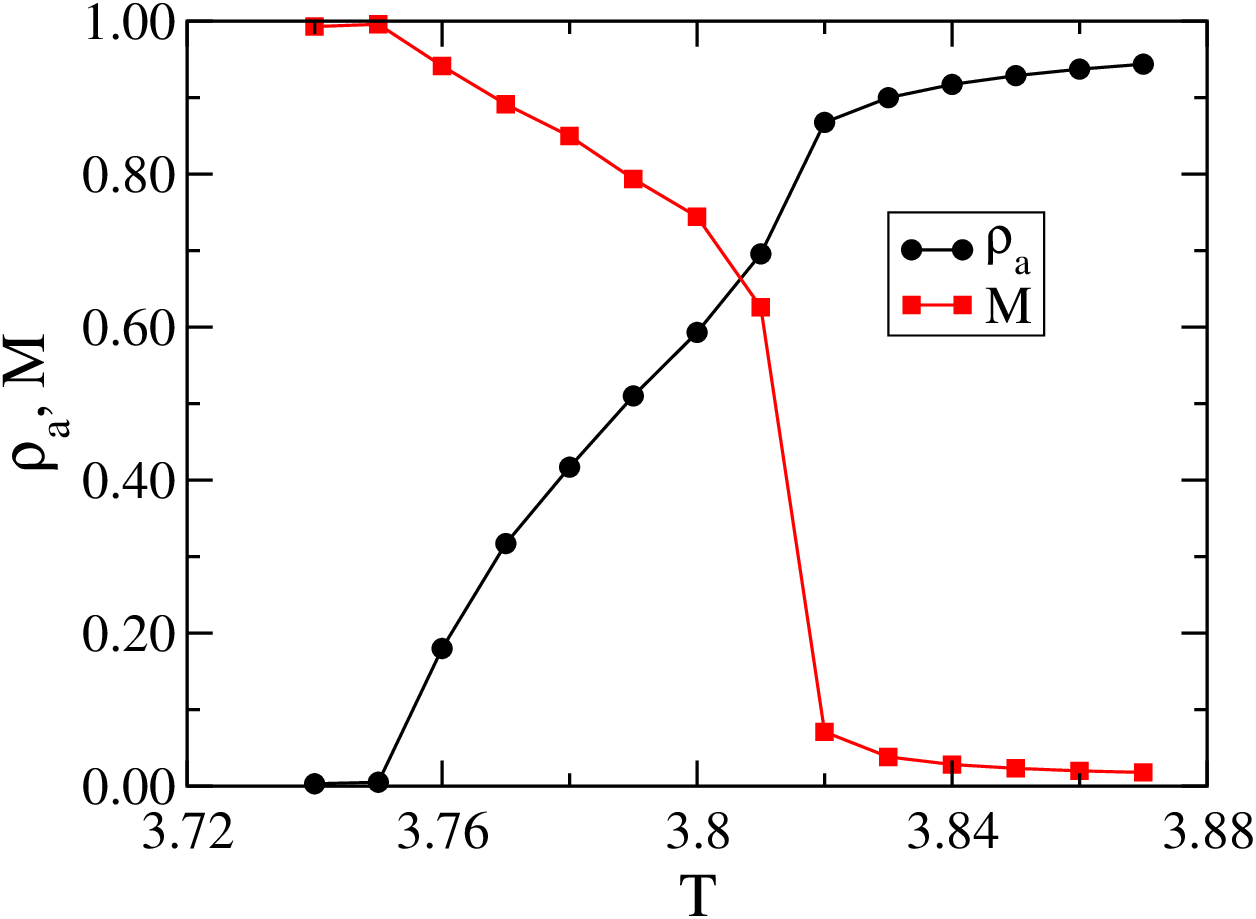}
\end{center}
\caption{
The steady-state density of active sites $\rho_a$ and the steady-state magnetization $M$ as a function of temperature for the LD model with $q=3$
and up to third-nearest-neighbor interactions. The behavior of the two order parameters indicate the presence of
two phase transitions that take place at different temperatures. The data, obtained for a system with
linear extent $L=400$, result from a time average (after having discarded the early time data) and an ensemble average over 150 independent runs.
Error bars are of the order of the symbol sizes.
}
\label{fig8}
\end{figure}

A first indication of the appearance of two phase transitions taking place at different temperatures can be seen in Figure \ref{fig8}
where we show for a system of linear extent $L=400$ the temperature dependence of the steady-state density of active sites $\rho_a$ and the 
steady-state magnetization $M$. The two quantities clearly approach zero at two different temperatures, thus exhibiting the same
behavior as the $q=2$ case \cite{Droz03a}.

\begin{figure}[ht]
\begin{center}
\includegraphics[angle=0,width=0.6\linewidth]{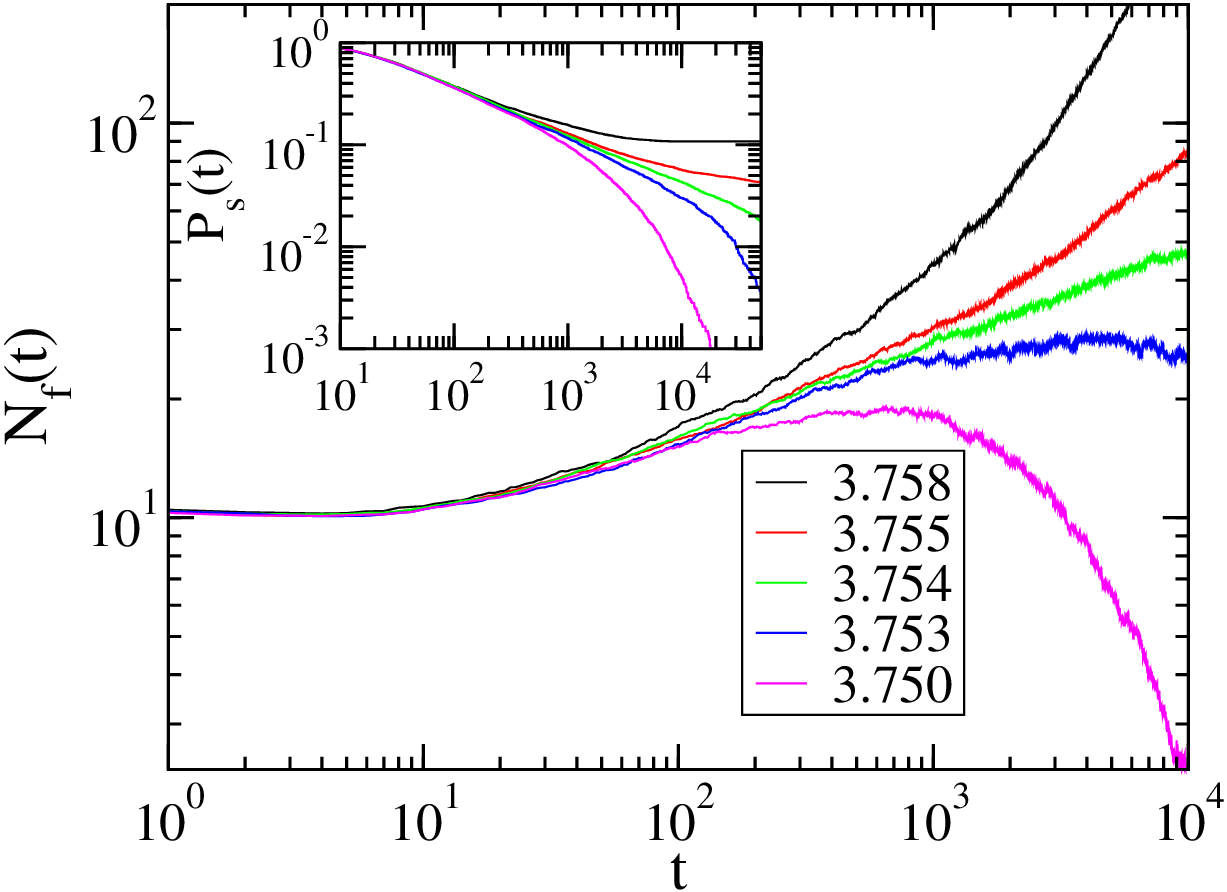}
\end{center}
\caption{
Number of flipped spins $N_f$ (main figure) and survival probability $P_s$ (inset) as a function of time 
for the LD model with $q=3$ and up to third-nearest-neighbor interactions. 
Both quantities display an algebraic behavior at the absorbing phase transition temperature $T_a = 3.754$.
The measured exponents $\eta = 0.238(10)$ for $N_f$ and $\delta=0.455(6)$ for $P_s$ agree with the values
of the Directed Percolation universality class \cite{Grassberger96,Voigt97}. The data, obtained for a system with $L=400$, result from averaging
over at least 10000 independent runs.
}   
\label{fig9}
\end{figure}

In order to better characterize the absorbing phase transition we show in Figure \ref{fig9} the time dependence of the number
of flipped spins $N_f(t)$ as well as that of the survival probability $P_s(t)$. As already described in Section 2, we prepare for
this measurement the system in one state, say 0, before assigning to a connected cluster of twelve spins randomly the values 1 or 2.
Having prepared the system in this way, we count at each time step the number of variables $\sigma_i$ with a value different from 0 and 
derive from this the two quantities shown in Figure \ref{fig9}.

We first note that both quantities display an algebraic behavior at the temperature $T=3.754$, whereas for temperatures slightly
above or below that value clear deviations from the power-law behavior are observed. This allows us to determine the
temperature of the absorbing phase transition to be $T_a = 3.754(1)$. Furthermore, the algebraic growth respectively decrease
of $N_f(t) \sim t^\eta$ respectively $P_s(t) \sim t^{-\delta}$ yields the exponent $\eta = 0.238(10)$ respectively $\delta=0.455(6)$. Comparing these
values with the known values $\eta = 0.230$ and $\delta=0.451$ for Directed Percolation \cite{Grassberger96,Voigt97},
we conclude that the absorbing phase transition encountered in the $q=3$ LD model with $N_i = 12$ indeed belongs to the
Directed Percolation universality class.

\begin{figure}[ht]
\begin{center}
\includegraphics[angle=0,width=0.6\linewidth]{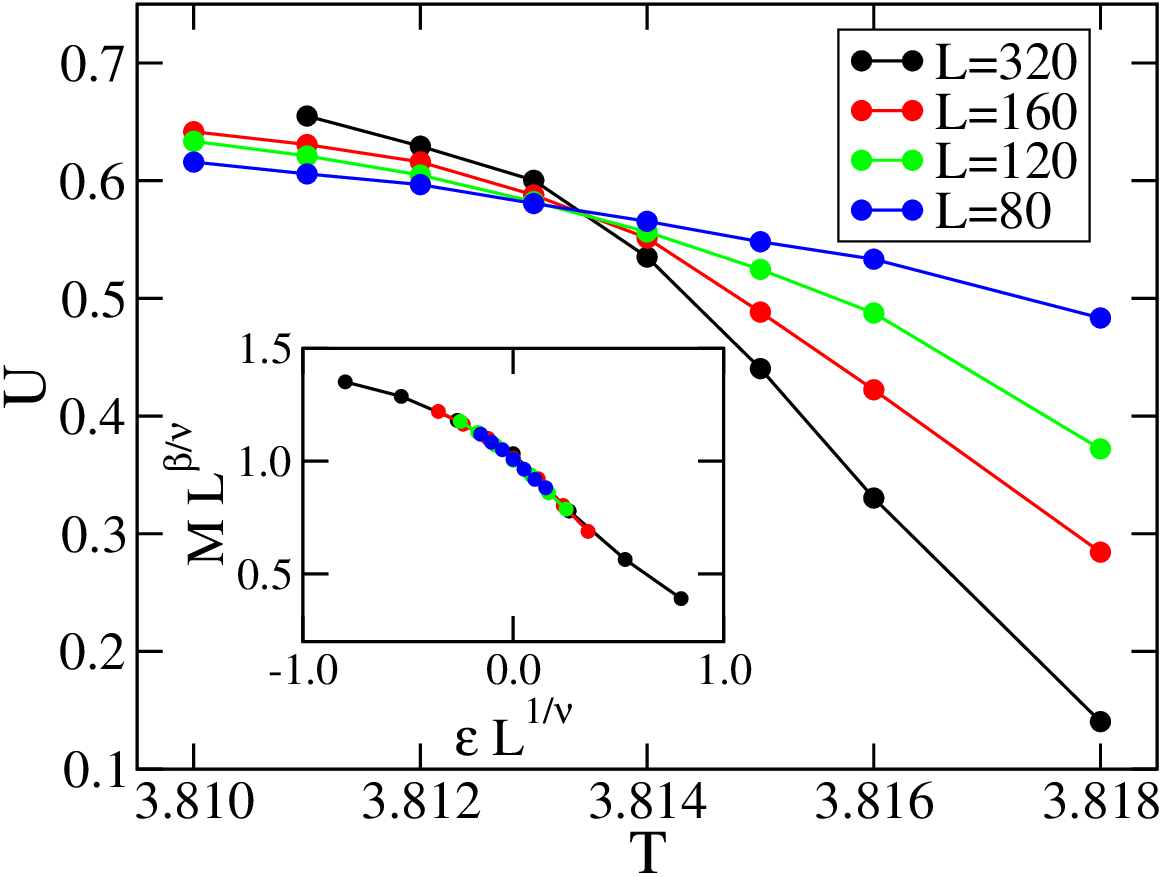}
\end{center}
\caption{
Main figure: Binder cumulant for the LD model with $q=3$ and up to third-nearest-neighbor interactions. Systems of different
sizes have been simulated. Inset: finite-size scaling of the magnetization $M$ when using the critical exponents of the
$q=3$ equilibrium Potts model. Error bars are smaller than the symbol sizes. The data result from averaging over 200
independent runs.
}   
\label{fig10}
\end{figure}

The critical temperature for the symmetry-breaking transition can be reliably determined by investigating the Binder cumulant 
for different system sizes, as shown in Figure \ref{fig10}. From the intersections of the different data sets we obtain
$T_c = 3.813(1)$, a value that is indeed larger than the temperature $T_a$ of the absorbing phase transition.
The universality class of this symmetry-breaking phase transition can be probed through finite-size scaling. As an example we show
in the inset of Figure \ref{fig10} the scaling plot for the magnetization where we plot $M L^{\beta/\nu}$ as a function of
$\varepsilon L^{1/\nu}$, with the reduced temperature $\varepsilon = \frac{T_c - T}{T_c}$. Inserting as $T_c$ the value we obtain from the crossing
of the Binder cumulant data as well as the known exponents $\beta = 1/9$ and $\nu = 5/6$ for the two-dimensional $q=3$ Potts
universality class results in the data collapse shown in the inset.

Finally, we also investigated the two-time autocorrelation and autoresponse functions
for this case (not shown). As for the $q=2$ LD model with up to third-nearest-neighbor
interactions, see Figure \ref{fig4}, these quantities suffer from sizeable finite-time corrections, due to the closeness of the
symmetry-breaking and the absorbing phase transitions. As a consequence one also has to go for $q=3$ to large waiting times in order to encounter
a clean data collapse with the aging exponents of the critical equilibrium three-state Potts model.

\section{Conclusion} \label{conclusion}
The aim of this paper has been to further elucidate the properties of a family of non-equilibrium Potts models with 
absorbing states that exhibit various scenarios
when changing the number of states $q$ of the spins or the range of the spin-spin interaction \cite{Lipowski02a,Lipowski02b,Droz03a,Droz03b}.
Most notably, in the presence of only nearest-neighbor interactions the model exhibits for the studied values of $q$ a temperature at which a symmetry-breaking
and an absorbing phase transition coincide, whereas for larger interaction ranges this common transition is split into a symmetry-breaking
transition at higher temperature and an absorbing phase transition at lower temperature. In our work we have characterized the different
phase transition through standard steady-state quantities as well as through various dynamic quantities, including two-time correlation
and response functions. In order to support some of our conclusions we have also presented results for other models such as the
generalized voter model introduced by Krause, B\"{o}ttcher, and Bornholdt \cite{Krause12} and the equilibrium two-dimensional
Potts model with $q > 4$.

Our main results are as follows:
\begin{itemize}
\item LD model with $q=2$ and only nearest-neighbor interactions.\\
In this case the system exhibits a single phase transition with coinciding symmetry-breaking and absorbing transitions. This phase transition
is known to belong to the generalized voter universality class. Both for the LD model and the KBB model, whose probabilities for spin flips
at the critical point are given by the transition probabilities of the linear voter model, we observe simple aging scaling for
the two-time autocorrelation and autoresponse functions. The scaling exponent and the scaling function are found to differ from
an analytical expression proposed in \cite{Hase10}.
\item LD model with $q=2$ and interactions with up to third nearest neighbors.\\
Extending the range of spin-spin interactions results in the splitting of the voter critical point into a symmetry-breaking phase transition,
expected to belong to the equilibrium Ising universality class, 
and an absorbing phase transition that take place at different temperatures. These two transition temperatures are very close, which 
can make it challenging to determine numerically the universality class of the symmetry-breaking transition when using steady-state
quantities. We discuss as an alternate approach
the investigation of aging quantities and observe strong finite-time corrections. As a result long waiting times are needed before
the expected aging scaling with Ising exponents is encountered.
\item LD model with $q=3$ and only nearest-neighbor interactions.\\
This case is characterized by a discontinuous phase transition at which both the symmetry-breaking order parameter as well as the absorbing order parameter
exhibit jumps. Two-time quantities display a behavior that is also observed at the first-order phase transition of the equilibrium Potts model
with larger values of $q$ (as verified for $q=7$ and $q=8$). 
A different, namely a simple aging scaling, behavior is encountered for the $q=5$ equilibrium Potts model. 
These differences are explained by the equilibrium correlation length which for $q=5$ is large compared to the system size, whereas
in the other studied cases it is smaller than the system size.
\item LD model with $q=3$ and interactions with up to third nearest neighbors.\\
As for the $q=2$ case with up to third-nearest-neighbor interactions we also find for $q=3$ two different phase transitions taking place at different, albeit
close, temperatures. The absorbing phase transition is shown to belong to the Directed Percolation universality class, whereas the
symmetry-breaking phase transition at higher temperature has the same critical exponents as the equilibrium $q=3$ Potts model in
two space dimensions.
\end{itemize}

In conclusion, our investigation of time-dependent quantities at the phase transitions encountered in a family of non-equilibrium Potts models 
with absorbing states clarifies the relaxation processes far from stationarity in cases where an order-disorder and an absorbing phase 
transition either take place simultaneously or are separated by only a very small change in temperature. The different scenarios
encountered in these models illustrate the intriguing properties, including the transient behavior far from
stationarity, that can be observed in systems with multiple critical points.

\section*{Acknowledgments} \label{acknowledgements}
Research was sponsored by the US Army Research Office and was accomplished under
Grant Number W911NF-17-1-0156. The views and conclusions contained in this document
are those of the authors and should not be interpreted as representing the official policies,
    either expressed or implied, of the Army Research Office or the US Government.

\appendix


\begin{thebibliography}{99}
\bibitem{footnote} Strictly speaking, $T$ is not a temperature in the non-equilibrium models discussed in this work. Still,
for convenience we will continue denoting the parameter $T$ as temperature in the following.
\end{thebibliography}
\end{document}